\documentclass[prd,nofootinbib,preprint,superscriptaddress,10pt]{revtex4-1}
\pdfoutput=1
\usepackage{amsmath,amssymb}
\usepackage{epsfig}
\usepackage{graphicx}
\usepackage[usenames,dvipsnames]{color}
\usepackage{subfigure}
\usepackage{slashed}
\usepackage[colorlinks,citecolor=blue]{hyperref}
\usepackage{color}

\usepackage{multirow}
\usepackage[normalem]{ulem}
\usepackage{xcolor}
\usepackage{soul}

\def\beq{\begin{equation}}
\def\eeq{\end{equation}}
\def\bea{\begin{eqnarray}}
\def\eea{\end{eqnarray}}
\DeclareUnicodeCharacter{2212}{-}
\begin{document}

\title{Imprint of inflationary gravitational waves and WIMP dark matter \\ in pulsar timing array data}

\author{Debasish Borah}
\email{dborah@iitg.ac.in}
\affiliation{Department of Physics, Indian Institute of Technology Guwahati, Assam 781039, India}
\author{Suruj Jyoti Das }
\email{suruj@iitg.ac.in}
\affiliation{Department of Physics, Indian Institute of Technology Guwahati, Assam 781039, India}
\author{Rome Samanta}
\email{samanta@fzu.cz}
\affiliation{CEICO, Institute of Physics of the Czech Academy of Sciences, Na Slovance 1999/2, 182 21 Prague 8, Czech Republic}

\begin{abstract}
Motivated by the recent release of new results from five different pulsar timing array (PTA) experiments claiming to have found compelling evidence for primordial gravitational waves (GW) at nano-Hz frequencies, we consider the prospects of generating such a signal from inflationary blue-tilted tensor power spectrum in a specific dark matter (DM) scenario dubbed as \textit{Miracle-less WIMP}. While \textit{Miracle-less WIMP}, due to insufficient interaction rate with the Standard Model (SM) bath gets thermally overproduced, inflationary blue-tilted gravitational waves (BGW) in compliance with PTA data, conflict cosmological observations if reheat temperature after inflation is sufficiently high. Both these issues are circumvented with late entropy dilution, bringing DM abundance within observational limits and creating a doubly-peaked feature in the BGW spectrum consistent with cosmological observations. The blue-tilted tail of the low-frequency peak can fit NANOGrav 15 yr data, while other parts of the spectrum are within reach of present and future GW experiments.




\end{abstract}

\maketitle

\section{Introduction}
\label{sec1}
Recently, five different pulsar timing array (PTA) experiments namely NANOGrav \cite{NANOGrav:2023gor}, European Pulsar Timing Array (EPTA) together with the first data release from Indian Pulsar Timing Array (InPTA) \cite{Antoniadis:2023ott}, PPTA \cite{Reardon:2023gzh} and Chinese Pulsar Timing Array (CPTA) \cite{Xu:2023wog} have released their latest findings hinting at significant evidence for stochastic gravitational waves (GW) background at nano-Hz frequencies supported by Hellings-Downs inter-pulsar correlations. While supermassive black hole binary (SMBHB) mergers can, in principle, generate such a signal, albeit with mild tension, plenty of scopes exist for exotic new physics \cite{NANOGrav:2023hvm}. Several follow-up papers have also studied the possible origin or implications of this observation from the point of view of DM \cite{Ghoshal:2023fhh, Shen:2023pan}, axions or axion-like particles \cite{Yang:2023aak, Guo:2023hyp}, SMBHB \cite{Ellis:2023dgf}, first-order phase transition \cite{Megias:2023kiy, Fujikura:2023lkn, Han:2023olf, Zu:2023olm}, primordial black holes \cite{Franciolini:2023pbf}, topological defects \cite{Kitajima:2023cek, Bai:2023cqj, Ellis:2023tsl, Wang:2023len}, inflation \cite{Vagnozzi:2023lwo} among others \cite{Li:2023yaj, Lambiase:2023pxd, Franciolini:2023wjm}.

In this paper, we propose a novel way of explaining the PTA data with GW generated from inflationary blue-tilted tensor power spectrum having unique correlation to a specific DM scenario known as \textit{Miracle-less WIMP} \cite{Borah:2022byb}. While weakly interacting massive particle (WIMP), the popular DM paradigm, has not shown up in direct search experiments yet, it may also indicate that DM perhaps interacts with the standard model (SM) bath more weakly. The \textit{Miracle-less WIMP} is a variant of the popular weakly interacting massive particle (WIMP) DM where DM-SM interactions fall short of the required WIMP DM criteria as suggested by null direct searches, but large enough to produce it in thermal equilibrium. While typical WIMP DM mass is restricted in between a few GeV \cite{Lee:1977ua} to a few hundred TeV \cite{Griest:1989wd}, \textit{Miracle-less WIMP} can have a much wider range of masses \cite{Borah:2022byb}. One natural way to achieve such a weaker cross-section is to consider a heavy mediator as a $U(1)$ gauge boson. The heavy gauge boson mediator arises from spontaneous $U(1)$ breaking, which also leads to the formation of stable cosmic strings (CS) \cite{Kibble:1976sj, Nielsen:1973cs,Vilenkin:1981bx,Turok:1984cn}. These CS can generate stochastic GW with a characteristic spectrum within reach of near-future GW detectors if the symmetry breaking scale is sufficiently high \cite{Borah:2022byb}. \footnote{Let's recall that 12.5 yr data from NANOGrav \cite{NANOGrav:2021flc} could be explained with stable cosmic string as the source of GW \cite{Blasi:2020mfx, Ellis:2020ena,Samanta:2020cdk,Borah:2022iym}. However, the 2023 data can not be fitted well with stable CS as the preferred slope pertinent to the new data \cite{NANOGrav:2023hvm} is inconsistent with the required amplitude.}

\begin{figure}
    \includegraphics[scale=.8]{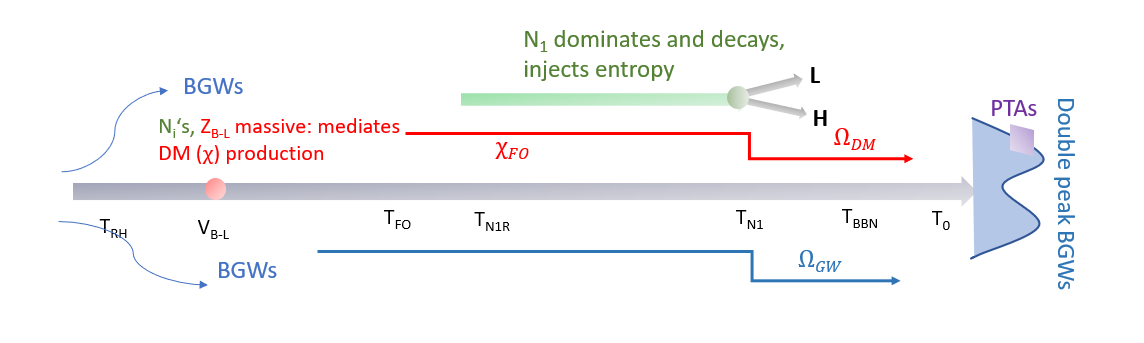}
    \caption{An illustrative timeline of the key phases discussed in this model. $T_{\rm RH}$ is the standard reheating temperature after inflation. At a scale $T\simeq v_{B-L}$, the $U(1)$ symmetry breaks. The DM freezes out relativistically at $T_{\rm FO}$. At $T_{N_{1}R}$, $N_1$ starts to dominate and at $T_{N_{1}}$, it decays to SM lepton and Higgs. $T_{\rm BBN}, T_0$ is the temperature corresponding to the BBN and present epochs respectively.  }\label{cartoon}
\end{figure}

While the particle physics setup we use here for illustration naturally exhibits GW due to stable CS \cite{Borah:2022byb}, we consider inflationary blue-tilted tensor fluctuations--blue-tilted gravitational waves (BGW), to be the primary source of GW to explain the recent PTA data (we shall see later that GWs from CS become irrelevant when this model is fitted to the PTA data with BGW). However, similar to \textit{Miracle-less WIMP} DM overclosing the universe due to thermal overproduction, BGW compatible with PTA data violates the bounds from big bang nucleosynthesis (BBN) and cosmic microwave background (CMB) on effective relativistic degrees of freedom $N_{\rm eff}$ for high reheating temperature after inflation, assuming the spectrum spans with a power-law at higher frequencies. Both of these issues can be tackled simultaneously by a common source of entropy dilution (a long-lived right-handed neutrino (RHN) $N_1$ in this model  \cite{Borah:2022byb}) in the early universe, which not only gives rise to consistency with observations but also leads to a GW spectrum that can explain the 2023 PTA data while being verifiable in future GW experiments at higher frequencies. This is in sharp contrast to the scenario without entropy dilution where the BGW explanation works only for reheating temperature as low as $\lesssim 10$ GeV \cite{Vagnozzi:2023lwo}. The entropy dilution required to satisfy the correct DM relic leads to a doubly-peaked feature in the BGW with the blue-tilted part of the low-frequency peak fitting NANOGrav 15 yr data at $2\sigma$ level. In addition, the DM mass and  peak frequencies of the BGW are uniquely correlated. The key phases in the early universe relevant to our discussion have been summarised in Fig.\ref{cartoon}.

Let's also mention that although typical slow-roll inflation models cannot produce such BGW, many models beyond slow-roll predict tensor blue tilt, e.g.,\cite{Gruzinov:2004ty,Kobayashi:2010cm,Endlich:2012pz,Cannone:2014uqa,Ricciardone:2016lym,Cai:2014uka,Fujita:2018ehq,Mishima:2019vlh}. Because the PTA experiments such as NANOGrav continue to prefer a positive slope of the GW spectrum, GW with tensor blue tilt are one of the most favorable candidates \cite{Vagnozzi:2020gtf,Bhattacharya:2020lhc,Kuroyanagi:2020sfw,Benetti:2021uea,Datta:2022tab}. Additionally, such GW not only exhibit testable characteristic spectral features at high frequencies but even for GW detection below nano-Hz frequencies \cite{DeRocco:2022irl,DeRocco:2023qae}, they are among only a few candidates with strong amplitude.

This paper is organised as follows. In section \ref{sec2}, we discuss the framework of Miracle-less WIMP DM followed by the details of inflationary blue-tilted GW in section \ref{sec3}. In section \ref{sec4}, we discuss the fit to PTA data with correlations to DM parameter space and finally conclude in section \ref{sec5}.

 \section{ Miracle-less WIMP Dark matter}
 \label{sec2}
To show a realistic scenario, we consider the example of gauged $U(1)_{B-L}$ model \cite{Davidson:1978pm, Mohapatra:1980qe, Marshak:1979fm, Masiero:1982fi, Mohapatra:1982xz, Buchmuller:1991ce}. As studied earlier \cite{Borah:2022byb}, a singlet Dirac fermion $\chi$ with $B-L$ charge $q_\chi$ can be the Miracle-less WIMP DM candidate, stabilized by a remnant $Z_2$ symmetry. DM overproduction and subsequent entropy dilution from heavy RHN, which are part of this model naturally, are dictated by the $U(1)_{B-L}$ parameters. However, our generic conclusion remains valid in any other setup as long as DM overproduction and required entropy dilution due to an early matter-dominated phase are realized. While DM and RHN are dominantly produced via interactions with $U(1)_{B-L}$ gauge boson $Z_{BL}$, the latter decays by virtue of tiny Yukawa couplings with leptons and Higgs.

For sufficiently heavy $Z_{BL}$, DM around or below the TeV ballpark freezes out from the bath while being relativistic. The DM relic density is then given by \cite{Kolb:1990vq}
\begin{align}
 \Omega_\chi h^2 & =2.745\times 10^{8}\times Y_\infty m_\chi,\label{eq:mainEQ}
\end{align}
where $Y_{\infty}=\frac{0.278}{g_{*s}(x_f)}\times \frac{3 g_\chi}{4}$ is the asymptotic comoving DM density with $g_{\chi}$ and $g_{*s}(x_f)$ being the DM internal degrees of freedom (dof) and entropy dof of the universe at DM freeze-out temperature $(T_{\rm FO} \equiv T_f= m_\chi/x_f)$ respectively\footnote{Here, although DM freezes out while being relativistic, the very early epoch of freeze-out and chosen DM mass (few MeVs to GeVs) lead to small free-streaming length (FSL) consistent with the cold dark matter paradigm and hence unconstrained from structure formation data. The FSL can be estimated as \cite{Merle:2013wta}
    \begin{eqnarray} \label{free-streaming}
    \lambda_{\rm FSL} = \int_{t_{\rm prod}}^{t_{\rm eq}}
\dfrac{\langle v\rangle}{a (t)} \,dt \simeq \frac{\sqrt{t_{\rm eq} t_{\rm nr}}} {a (t_{\rm eq})} \left( 5 + \text{ln} \left(\frac{t_{\rm eq}}{t_{\rm nr}}\right)\right)\,.
\nonumber 
\end{eqnarray}
Here, $a$ denotes the scale factor, $t_{\rm prod}$  is the epoch when maximum production of dark matter occurs,  $t_{\rm eq}\sim 2\times 10^{12} $ s, is the epoch of matter-radiation equality after which the structure formation starts, and  $\langle v\rangle$ is the average velocity of DM. $t_{\rm nr}$ indicates the epoch when DM becomes non-relativistic. For DM mass $m_{\chi}\sim$ MeV-GeV, we find $\lambda_{\rm FSL}\sim$ $0.001$-$10^{-7}\,\text{Mpc}$, which is quite small to be constrained from structure formation data.}. We also consider $g_{*s}(x_f)=106.75$ for the SM entropy dof as the freeze-out occurs well above the electroweak scale. If $\chi$ leads to overabundance, the required entropy dilution (facilitated by $N_1$ decay) factor $S = \Omega_\chi h^2/0.12$ can be approximated as \cite{Scherrer:1984fd},
\begin{align}
 S\simeq \left[2.95\times \left(\frac{2\pi^2 {g}_{*}(T_{N_1})}{45}\right)^{1/3}\frac{(Y_{N_1} M_{N_1})^{4/3}}{(\Gamma_{N_1} M_P)^{2/3}}\right]^{3/4},\label{eq:dilutionF}
\end{align}
where ${g}_{*}(T_{N_1})$ is the number of relativistic dof during $N_1$ decay at $T=T_{N_1}$. We note that $g_* \sim g_{*s}$ for the region of our interest. The parameter $Y_{N_1}$ is the freeze-out number density of $N_1$ considering relativistic freeze out 
\begin{align}
 Y_{N_1}=\frac{g_{N_1}}{2}\frac{135\,\zeta(3)}{4\pi^4 g_{*}^{\rm fo}}.\label{eq:N1abun}
\end{align}

Assuming instantaneous decay ($ \Gamma_{N_1} M_P=1.66 \sqrt{{g}_{*}(T_{N_1})} T_{N_1}^2$), we find
 \begin{align}
  T_{N_1} \simeq 3.104\times 10^{-10}\left(\frac{M_{N_1}}{m_\chi}\right) \rm GeV.\label{eq:anaEntropy}
 \end{align}
Eq.(\ref{eq:anaEntropy}) shows that for a fixed $M_{N_1}$, larger $m_\chi$ corresponds to smaller $T_{N_1}$, which means $N_1$ dominates for a longer period and produces larger entropy. This result can be intuitively inferred from Eq.(\ref{eq:mainEQ})--heavier $\chi$ corresponds to a large initial number density, requiring stronger dilution. 

\begin{figure}
    \includegraphics[scale=.33]{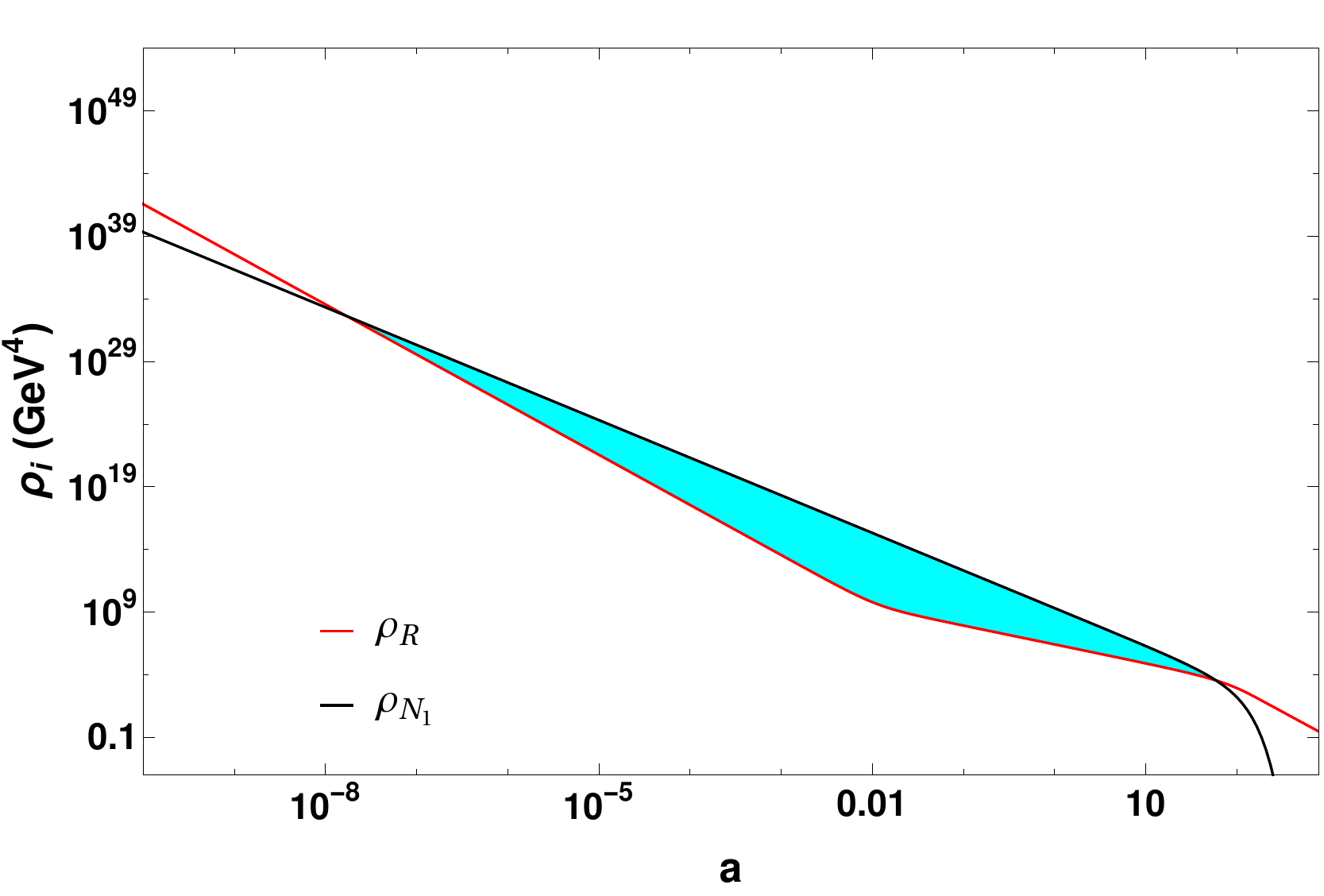}
    \caption{Evolution of the radiation energy density (red contour) and the energy density of $N_1$ (black contour), for the BP2 (cf. Fig. \ref{fig1}). We choose the initial value of scale factor as $a_{\rm in}= \frac{1 \text{GeV}}{T_{\rm in}}$, where the initial temperature $T_{\rm in}\sim v_{\rm B-L}$.}\label{fig:evo}
\end{figure}

In order to track the evolution of the energy components of radiation, DM and the diluter $N_1$, we solve a set of coupled Boltzmann equations given by\footnote{Note that $N_1$ can decay into SM particles due to the Dirac Yukawa couplings given by $\sum_{\substack{\\\alpha=e, \mu, \tau}}Y_D^{\alpha 1}~\overline{l_{L}^{\alpha}}\tilde{H_{\rm S}}N_{1}\nonumber$, where ${H_{\rm S}}$ denotes the SM Higgs. The decay rate of $N_1$ can be written in terms of the Dirac Yukawa coupling as $\Gamma_{N_1}\simeq\frac{1}{16\pi}\sum_{\alpha}\big|Y_{D}^{\alpha 1}\big|^2 M_{N_{1}}$, where $Y_{D}^{\alpha 1}$ follows the Casas Ibarra parametrisation \cite{Casas:2001sr, Borah:2021inn, Borah:2022byb}. The required decay rate can be realised by choosing tiny Dirac Yukawa couplings which in turn predicts vanishingly small lightest active neutrino mass (shown in \cite{Borah:2022byb}). This can be refuted by tritium beta decay experiments such as KATRIN \cite{KATRIN:2019yun}, sensitive to absolute neutrino mass in the eV ballpark.} 
\begin{align}
& \frac{d E_{\chi}}{da}=\frac{\langle\sigma v\rangle_{\chi}}{Ha^{4}}\left((E_\chi^{\rm eq})^{2}-E_{\chi}^{2}\right) \, , \label{eq:bol1}\\
&\frac{d E_{N_{1}}}{da}=\frac{\langle\sigma v\rangle_{N_1}}{Ha^{4}}\left((E_{N_{1}}^{\rm eq})^{2}-E_{N_{1}}^{2}\right)-\frac{\Gamma_{N_1}}{Ha}E_{N_{1}} \, , \label{eq:bol2}\\
&\frac{dT}{da}=\left(1+\frac{T}{3 g_{*s}}\frac{dg_{*s}}{dT}\right)^{-1}\left[-\frac{T}{a}+\frac{\Gamma_{N_1}M_{N_1}}{3 H ~s~ a^4}E_{N_1}\right],  
\label{eq:Boltz}
\end{align}
\noindent where $s=\frac{2\pi^2}{45}g_{*s}T^3$ is the entropy density with $T$ indicating the temperature of radiation bath. The quantities $E_{\chi,N_{1}}$ represent the co-moving number densities, defined as $E_{\chi,N_{1}}=n_{\chi,N_{1}}a^3$ where $n_{\chi,{N_1}}$ are the usual number densities. $E_{\chi,N_{1}}^{\rm eq}$ denotes the equilibrium comoving number densities. In Fig. \ref{fig:evo}, we show the evolution of the energy densities of radiation ($\rho_R=\frac{\pi^2}{30}g_{*}T^4 $) and the diluter $N_1$ ($\rho_{N_1}$), which eventually comes to dominate the energy density (indicated by the shaded region). From the numerical solution, we find that the temperature $T_{N_1}$ at which $N_1$ domination ends agrees quite well upto a factor of $\sim2$, with the analytical estimate given by Eq. \eqref{eq:anaEntropy}, which assumes an instantaneous decay. 

\section{BGWs from inflation}
\label{sec3}
The following perturbed FLRW line element describes GW:
\bea
ds^2=a(\tau)\left[-d\tau^2+(\delta_{ij}+h_{ij})dx^idx^j)\right],
\eea
where $\tau$ and $a(\tau)$ are the conformal time and scale factor respectively. The transverse traceless ($\partial_ih^{ij}=0$, $\delta^{ij}h_{ij}=0$) part of  $h_{ij}$ represents the GW. After the Fourier space decomposition of $h_{ij}$ and solving GW propagation equation in Fourier space, the energy density of the GW is computed as \cite{WMAP:2006rnx}
\bea
\rho_{\rm GW}=\frac{1}{32\pi G}\int\frac{dk}{k}\left(\frac{k}{a}\right)^2T_T^2(\tau, k)P_T(k),\label{gw1}
\eea
where $T_T^2(\tau, k)=|h_k(\tau)|^2/|h_k(\tau_i)|^2$ is a transfer function  with $\tau_i$ as the initial conformal time, and $k=2\pi f$ with $f$ being the present frequency. The quantity $P_T(k)=\frac{k^3}{\pi^2}|h_k(\tau_i)|^2$ characterizes the primordial power spectrum and relates to the inflation models with specific forms, which, generally, is parametrized as a power-law:
\bea
P_T(k)=r A_s(k_*)\left(\frac{k}{k_*}\right)^{n_T},
\eea
where $r\lesssim 0.06$ \cite{BICEP2:2018kqh} is the tensor-to-scalar-ratio,  $A_s \simeq 2\times 10^{-9}$ is the scalar perturbation amplitude determined at the pivot scale $k_*=0.01\rm  Mpc^{-1}$. We shall treat the tensor-spectral index $n_T$ as constant plus blue-tilted ($n_T>0$). Recall that the single field slow-roll inflation models correspond to the consistency relation: $n_T=-r/8$ \cite{Liddle:1993fq}, i.e., the spectral index is mildly red-tilted ($n_T\lesssim 0$). The GW energy density pertinent to detection purposes is expressed as 
\bea
\Omega_{\rm GW}(k)=\frac{k}{\rho_c}\frac{d\rho_{\rm GW}}{dk},
\eea
where $\rho_c=3H_0^2/8\pi G$ with $H_0\simeq 2.2 \times 10^{-4}~\rm Mpc^{-1}$ being the Hubble constant. From Eq.\eqref{gw1}, the $\Omega_{\rm GW}(k)$ can be derived as 
\bea
\Omega_{\rm GW}(k)=\frac{1}{12H_0^2}\left(\frac{k}{a_0}\right)^2T_T^2(\tau_0,k)P_T(k),\label{GWeq}
\eea
where $\tau_0=1.4\times 10^4 {\rm ~Mpc}$.
The transfer function has been computed very accurately in literature \cite{Seto:2003kc,Boyle:2005se,Nakayama:2008wy,Kuroyanagi:2008ye,Nakayama:2009ce,Kuroyanagi:2014nba}. In presence of an intermediate matter domination $T_T^2(\tau_0,k)$ is calculated as \cite{Nakayama:2009ce,Kuroyanagi:2014nba}
\bea
T_T^2(\tau_0,k)=F(k)T_1^2(\zeta_{\rm eq})T_2^2(\zeta_{N_1})T_3^2(\zeta_{N_1 R})T_2^2(\zeta_{R}),
\eea
where $F(k)$ is given by
\bea
F(k)=\Omega_m^2\left( \frac{g_*(T_{k,\rm in})}{g_{*0}}\right)\left(\frac{g_{*s0}}{g_{*s}(T_{k,\rm in})}\right)^{4/3}\left(\frac{3j_1(k\tau_0)}{k\tau_0}\right)^2.\label{fuk}
\eea
 
In Eq.\eqref{fuk}, $j_1(k\tau_0)$ is the spherical Bessel function, $\Omega_m=0.31$, $g_{*0}=3.36$, $g_{*0s}=3.91$ and an approximate form of the scale-dependent $g_*$ can be found in \cite{Kuroyanagi:2014nba} 
The individual transfer functions read
\bea
T_1^2(\zeta)=1+1.57\zeta+ 3.42 \zeta^2,\\
T_2^2(\zeta)=\left(1-0.22\zeta^{1.5}+0.65\zeta^2 \right)^{-1},\\
T_3^2(\zeta)=1+0.59\zeta+0.65 \zeta^2,
\eea
where $\zeta_i\equiv k/k_i$, with the modes $k_i$'s in the units of ${\rm Mpc}^{-1}$ given by
\bea
k_{\rm eq}=7.1\times 10^{-2}\Omega_m h^2
\eea
\bea
k_{N_1}=1.7\times 10^{14}\left(\frac{g_{*s}(T_{N_1})}{106.75}\right)^{1/6} \left(\frac{T_{N_1}}{10^7 \rm GeV}\right),\label{1p}
\eea
\bea
k_{N_1 R}=1.7\times 10^{14} S^{2/3}\left(\frac{g_{*s}(T_{N_1})}{106.75}\right)^{1/6}\left(\frac{T_{N_1}}{10^7 \rm GeV}\right)
\eea
and 
\bea
k_{R}=1.7\times 10^{14}S^{-1/3}\left(\frac{g_{*s}(T_{\rm RH})}{106.75}\right)^{1/6}\left(\frac{T_{\rm RH}}{10^7 \rm GeV}\right)\label{2p}
\eea
that cross the horizon at standard matter-radiation equality temperature $T_{\rm eq}$, at $T_{N_1}$ when $N_1$ decays, at $T_{N_1R}$ when $N_1$ starts to dominate the energy density and at $T_{\rm RH}$ when the universe first reheat after inflation, respectively. Two major constraints on BGW arise from observed $N_{\rm eff}$ and LIGO bound on stochastic GW. The BBN constraint is given by \cite{Peimbert:2016bdg}
\bea
\int_{f_{\rm low}}^{f_{\rm high}} f^{-1}df \Omega_{\rm GW}(f)h^2\lesssim 5.6\times 10^{-6}\Delta N_{\rm eff},
\eea
with $\Delta N_{\rm eff}\lesssim 0.2$. The frequency $f_{\rm low}$ corresponds to the mode entering the horizon at the BBN epoch, which can be taken as $f_{\rm low}\simeq 10^{-10}$ Hz. On the other hand, we take $f_{\rm high}\simeq 10^5$ Hz, which is sufficient for numerical computation as the spectrum falls and the integration saturates at higher frequencies. We consider the LIGO bound in a much simpler way. We discard GW with amplitude more than $2.2\times 10^{-9}$ at $f_{\rm LIGO}=25$ Hz \cite{KAGRA:2021kbb}. Given the above equations and constraints, we now compute the GW spectrum for a few benchmark values.

\section{Fit to the PTA data and discussion}
\label{sec4}

\begin{figure*}
    \includegraphics[scale=0.6]{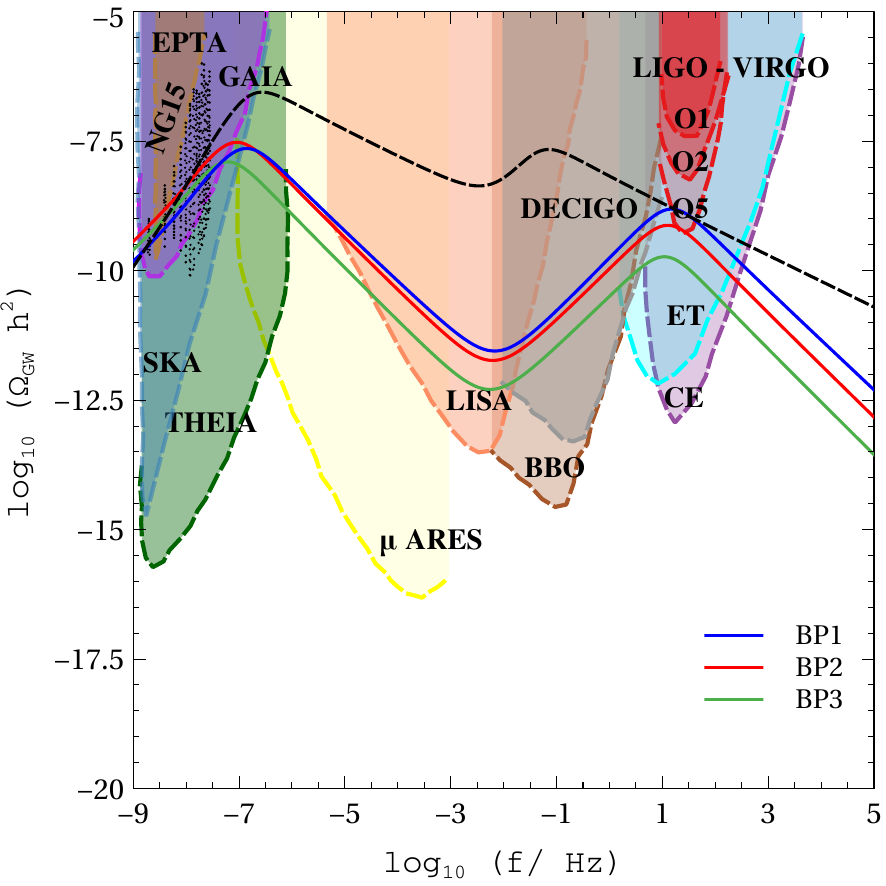} 
    \caption{Doubly-peaked GW spectrum for the benchmark points (BP1: blue, BP2: red, BP3: green). The black-dashed line is produced with $n_T=1.5$, $r=2 \times 10^{-7}$, $T_{\rm RH}=3 \times 10^8$ GeV, $T_{N_1}=5$ GeV and $S= 10^{7}$. This one is a `fake' spectrum. Meaning, despite the spectrum being consistent (actually gives a better fit than the BPs, see the left panel of Fig. \ref{fig1}) with the PTA data, the corresponding $T_{\rm RH}$ is incompatible with the present dark matter model.}
    \label{fig1a}
\end{figure*}

\begin{figure*}
\includegraphics[scale=0.55]{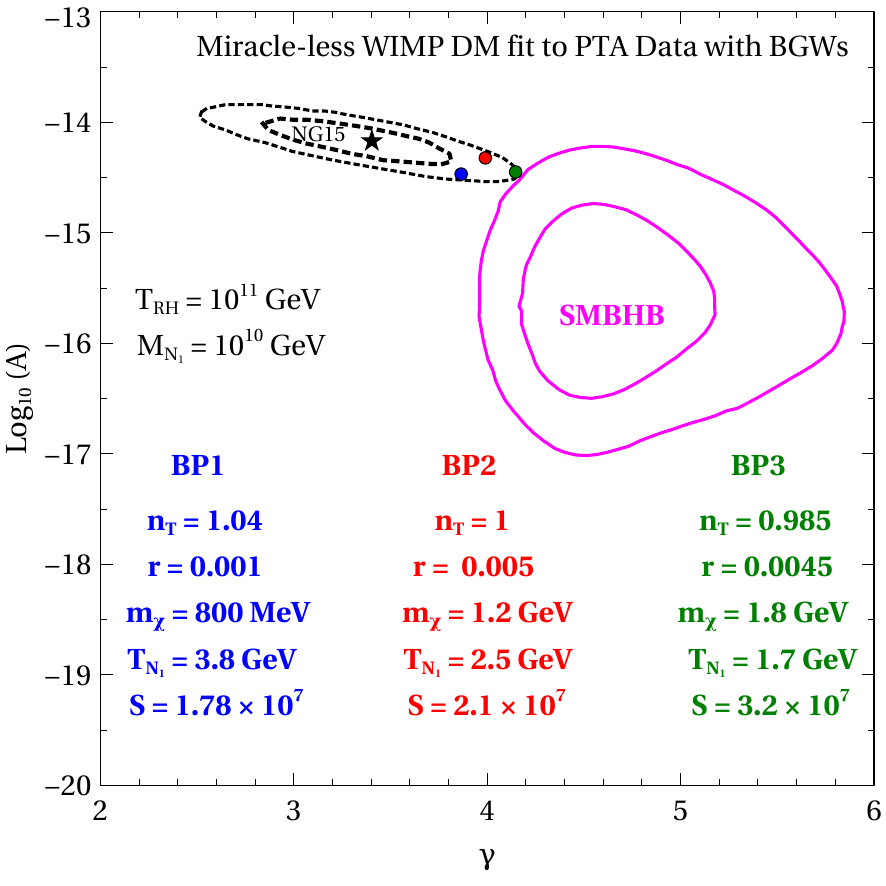}
     \includegraphics[scale=0.55]{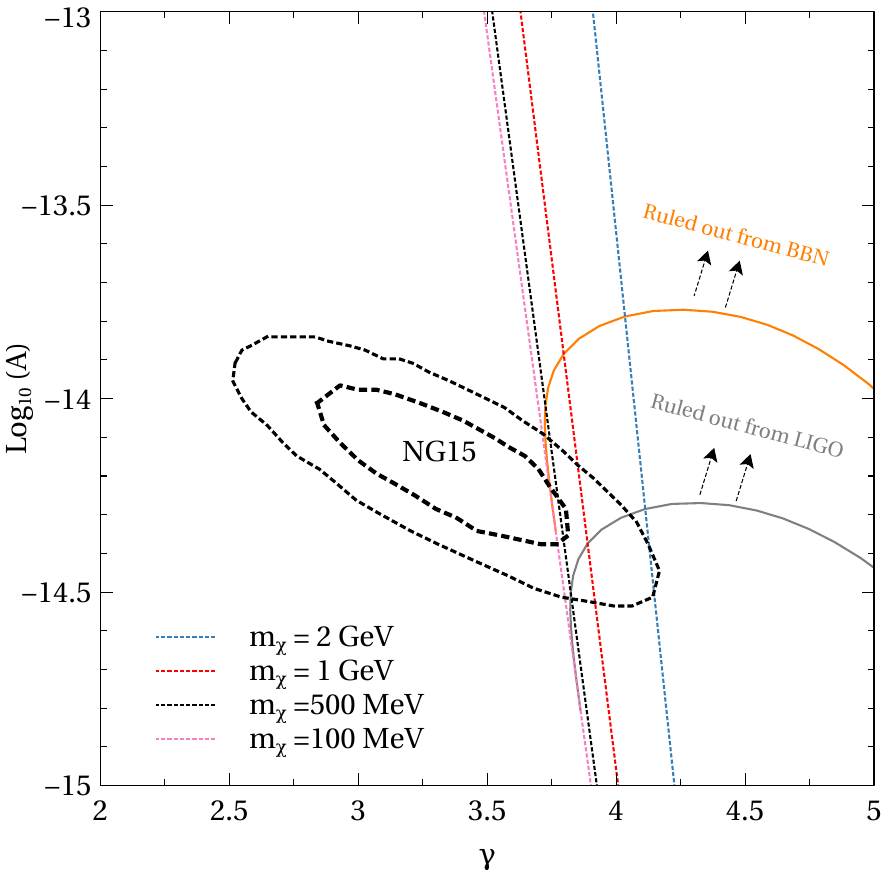}
    \caption{Left panel: Fit to the NANOGrav-2023 data for the same benchmark points, along with the results predicted for SMBHB. The `$\bigstar'$ corresponds to the spectrum shown with the black-dashed line in Fig.\ref{fig1a}. Right panel: Fit to the NANOGrav-2023 data for different DM masses, fixing $r= 0.001$, $T_{\rm RH} = 10^{11}$ GeV. The tensor spectral index $n_T$ varies along the straight lines. The region above the orange and gray contours are ruled out from BBN and LIGO bounds, respectively (see text).}
    \label{fig1}
\end{figure*}

First, notice that barring $n_T$ and $r$, the key quantities to evaluate the spectrum are $T_{N_1}$, $S$ and $T_{\rm RH}$. By construction, in this model, $T_{\rm RH}$ should be large--at least $\mathcal{O}(v_{B-L})$ (because we need heavy $Z_{BL}$ with mass $\sim v_{B-L}$ for weaker DM interaction cross-section and in addition, the DM plus the $N_1$ number densities are computed in first radiation domination after the universe reheats at $T_{\rm RH}$). However, BGW with large $n_T$ are incompatible with high $T_{\rm RH}$ (the amplitude saturates BBN and LIGO bounds) unless there is large entropy dilution. The Miracle-less WIMP scenario naturally exhibits intermediate matter domination by $N_1$, leading to large entropy production, which brings overproduced DM density within observed limits. Such large entropy production also suppresses the overall GW spectrum, plus depending on $T_{N_1}$, it creates another peak in the overall spectrum. Note from Eq.\eqref{eq:dilutionF} and Eq.\eqref{eq:anaEntropy} that two free parameters of the model $M_{N_1}$ and $m_\chi$ enter in the computation of GW through $S$ and $T_{N_1}$ and determine its spectral features. In Fig.\ref{fig1a}, we show the corresponding spectrum for three benchmarks points with $T_{\rm RH}=10^{\rm 11}$ GeV (BP1: blue, BP2: red, BP3: green). The benchmarks are chosen to fit the recent NANOGrav results to some extent, as we will discuss shortly. In principle, this model allows higher $T_{\rm RH}$, but one needs substantial entropy production to surpass the LIGO bound. However, in that case, the low-frequency GW amplitudes also get suppressed. Therefore,  even though the spectral index is compatible with NANOGrav, the overall amplitude falls below the reported range. 


\begin{figure*}
    \includegraphics[scale=0.33]{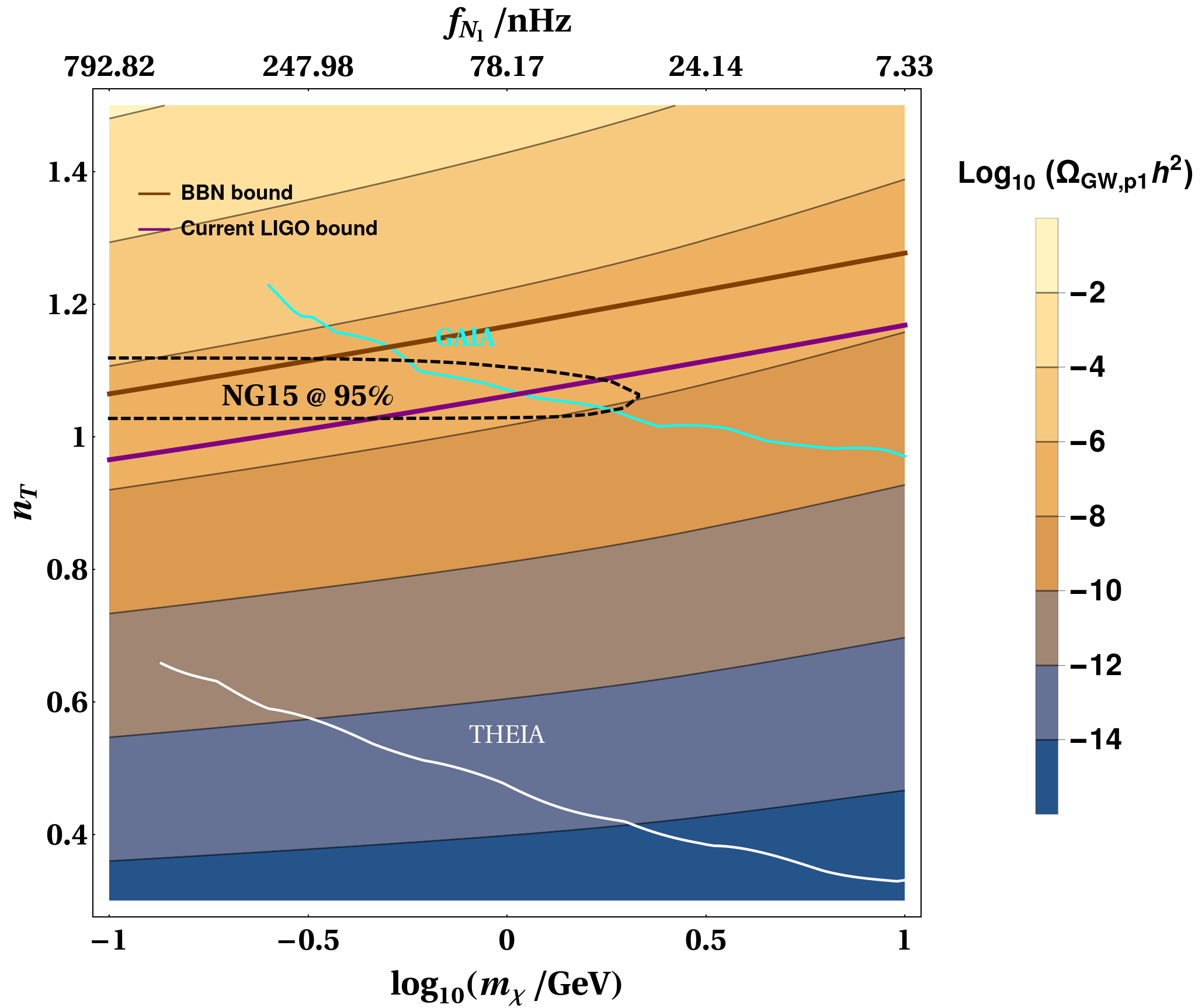} \includegraphics[scale=0.33]{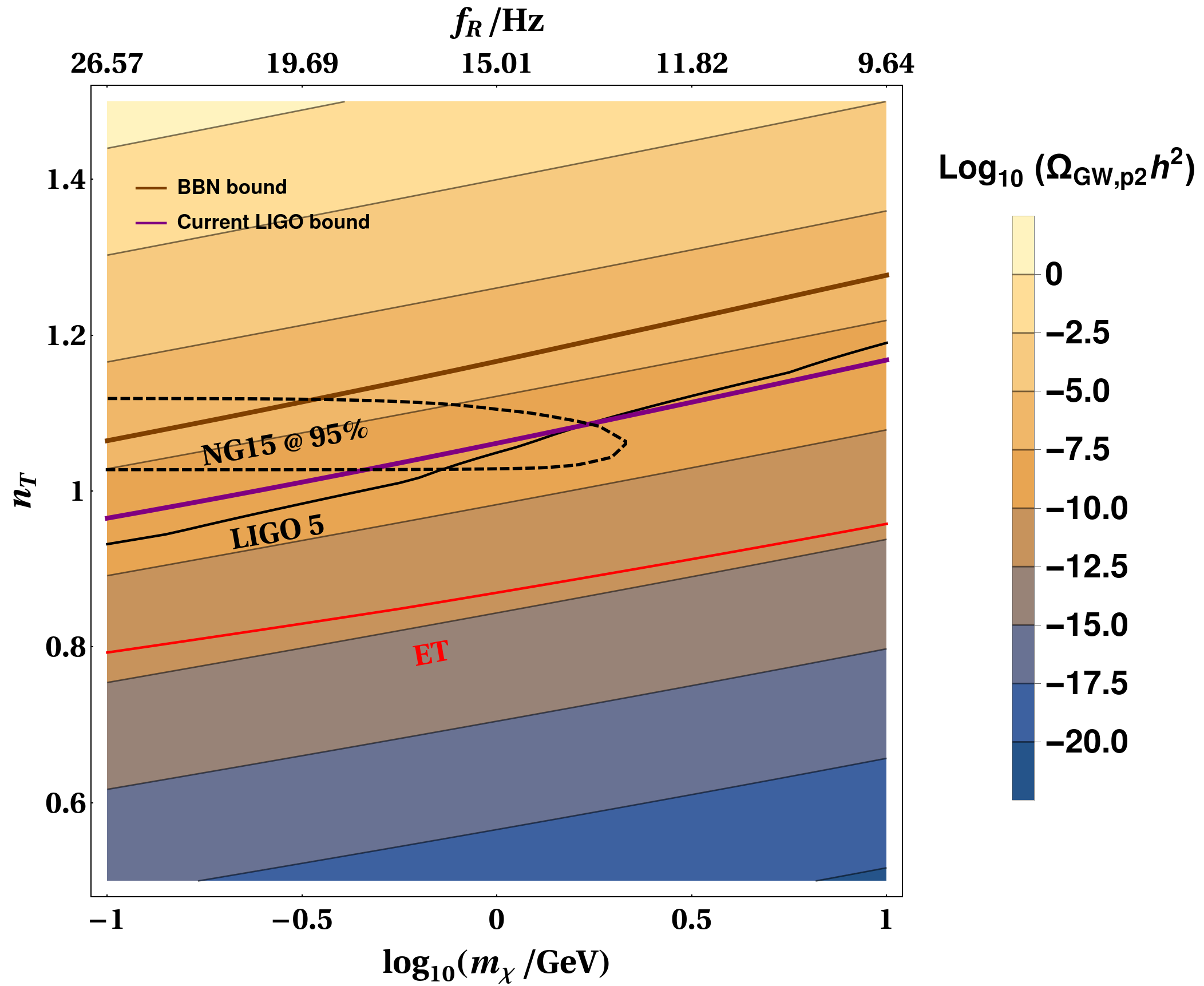}
    \caption{Contours of the $1^{\rm st}$ peak $\Omega_{\rm GW,p1}h^2$ (left panel), $2^{\rm nd}$ peak $\Omega_{\rm GW,p2}h^2$ (right panel) on the $m_{\chi}-n_T$ plane, along with the sensitivities of different GW experiments and current LIGO and BBO bounds. Here, we fix $r= 0.001$, $T_{\rm RH} = 10^{11}$ GeV. The values of $n_T$ and $m_\chi$ inside the black-dashed contour fit NANOGrav 2023 data at 2$\sigma$ level.}
    \label{figcontour}
\end{figure*}

We fit the NANOGrav-2023 data with a power-law signal represented by the characteristic strain 
\bea
h_c(f)=A\left(\frac{f}{f_{\rm yr}}\right)^{(3-\gamma)/2}, 
\eea
where $A$ and $\gamma$ are the strain amplitude and the timing-residual cross-power spectral index ($\gamma$ = 13/3 for super-massive black hole mergers) respectively, and $f_{\rm yr}={\rm yr}^{-1}$. The normalised GW energy density is expressed in terms of strain as
\bea
\Omega_{\rm GW}(f)=\frac{2\pi^2}{3 H_0^2}f^2h_c(f)^2=\Omega_{yr}\left(\frac{f}{f_{\rm yr}}\right)^{5-\gamma}, \label{pl}
\eea
where $\Omega_{\rm yr}=\frac{2\pi^2}{3 H_0^2}A^2f_{\rm yr}^2$. We fit Eq.\eqref{GWeq} to Eq.\eqref{pl} within the frequency range $f\in \left[2\times 10^{-9}, 3\times 10^{-8}\right]$ for the chosen benchmarks, extract $A$ and $\gamma$ from the fit and project it on NANOGrav $95\%$ and $68\%$ contours as shown in Fig.\ref{fig1} (left panel). The BPs lie close to the edge of the $95\%$ contour because the spectral index $\gamma\simeq 5-n_T$. Therefore, for $n_T\sim 1$, the benchmarks lie close to $\gamma \sim 4$. In principle, a large $n_T$ and small $r$ can provide a better fit (As shown with the black-dashed curve for $r=2\times 10^{-7}$, $n_T=1.5$, and $T_{\rm RH}=3\times 10^8$ GeV). Although it might seem that the fit improves for larger values of $n_T$, first, we note that $n_T$ as large as, e.g., 1.8 (because the NANOGrav best-fit values are $\gamma=3.2\pm 0.6$), is extremely difficult to obtain while being consistent with the constraints on other inflationary observables \cite{Wang:2014kqa}. Second, even leaving aside the discussion regarding the origin of large $n_T$, this model does not allow such a choice. This is because the benchmarks chosen to generate the black-dashed spectrum in Fig.\ref{fig1a} (or corresponding $\bigstar$ in Fig.\ref{fig1}) are inconsistent with the condition $v_{B-L}\lesssim T_{\rm RH}$, which is essential to produce Miracle-less WIMP. We, therefore, conclude that the recent NANOGrav data (or, more generally, the PTA data, which are in good agreement) fit well with inflationary GW and Miracle-less WIMP dark matter within $95\%$ CL. Therefore, this model, alongside providing a natural explanation of null results in direct searches, serves, perhaps, as the first WIMP model to bring amplitudes of the inflationary GW down to the level of PTAs despite a large $T_{\rm RH}$. In addition, this model predicts WIMP DM in the MeV-GeV ballpark, the tensor-to-scalar ratio $r\sim 10^{-3}$ (a fit to the NANOGrav 2023 data for this value has been presented in the right panel of Fig.\ref{fig1}) and another high-frequency peak that can be constrained by the interferometers such as LIGO (cf. Fig.\ref{fig1a}), making it a unique DM scenario that can be tested with PTA-LIGO complementarity.  \\

We conclude with the following remarks: \\

$\bullet$ Note that we can perform a simple power-law fit in this model because the first peak of the BGWs can always be made outside the highest frequency bin of NANOGrav (we take it to be $f_h\sim$ 1yr$^{-1}$) while being consistent with the reported range of the spectral index $\gamma$ (see Fig.\ref{fig1}, left). Therefore, any statistical comparison, e.g., with SMBHBs, should be made from the $A-\gamma$ plot, comparing the $\sigma$ contours. Unfortunately, most cosmological sources do not fit the NANOGrav new data well with a power-law. This is because, despite the possibility of having $f_{\rm peak}>f_h$, those sources do not have the required power in the frequency. Notable examples are the first-order phase transition and domain walls with constant tension. The infrared tail of these sources: $\Omega_{GW}(f<f_{\rm peak})\sim f^3$ is fixed by causality. Therefore, one obtains $\gamma=5-3=2$--far outside the $2\sigma$ contour. Nonetheless, they comply with data with a Bayesian fit considering $f_{\rm peak}<f_h$. In the NANOGrav catalogue \cite{NANOGrav:2023hvm}, most cosmological sources have been compared with SMBHBs by comparing their marginal likelihoods with a Bayes factor.\\

$\bullet$ To suppress inflationary GW and bring them to the level of PTAs, generally one needs heavy $N_1$ ($M_{N_1} \sim 10^{10}$ GeV) so that one produces large entropy according to Eq.\eqref{eq:dilutionF}. Therefore, dark matter cannot be arbitrarily heavy. Otherwise, it would require an extremely late time decay of $N_1$ (cf. Eq.\eqref{eq:anaEntropy}), which contradicts  BBN predictions.\\

$\bullet$ In this setup, the DM mass $m_\chi$ and frequencies $f_{N_1}$ (first peak), $f_{N_{1},R}$ (dip in the middle), and $f_R$ (second peak) are correlated, see Eq.(\ref{1p}-\ref{2p}) along with Eq.(\ref{eq:dilutionF}) and Eq.(\ref{eq:anaEntropy}). Experimental prospects of such a correlation, e.g., $m_\chi \leftrightarrow f_{N_1}$ and $m_\chi \leftrightarrow f_R$ have been summarised in Fig.\ref{figcontour}. Considering a fixed value of $r$, we have shown the contours of the $1^{\rm st}$ and the $2^{\rm nd}$ peak in the $m_{\chi}- n_T$ plane, along with several GW experiments which can probe them. On the other hand, the values of $m_{\chi}$ and $n_T$ inside the black-dashed contour fit the current NANOGrav data at 2$\sigma$ level. To understand this claim clearly, let's have a look at the right panel of Fig. \ref{fig1}. Note that along the contours, $n_T$ varies while $r$ is kept fixed. Considering for instance DM mass $m_{\chi} = 100$ MeV (pink-dashed line), the allowed range of $n_T$ is between the two points where it intersects the $2 \sigma$ contour. Moving towards the right, the DM mass increases, and for $m_{\chi} = 2$ GeV (blue-dashed line) the range of $n_T$ values turns out to be very small. For even larger values of DM mass, there exist no $n_T$ values that can fit the NANOGrav data. This explains the pattern observed in Fig. \ref{figcontour}.  Thus, our scenario predicts very specific values of DM mass which can fit the current NANOGrav data, while the other parts of the spectrum being testable at several future GW experiments. \\  

$\bullet$ The scenario also predicts CS that radiate GW \cite{Borah:2022byb}. Therefore one expects further spectral distortion as in \cite{Datta:2022tab}. However, for $v_{B-L}\sim T_{\rm RH}\sim 10^{11}$ GeV, the amplitude of the cosmic string radiated GW would be much smaller (max. $\Omega_{\rm GW}^{CS} \sim 10^{-13}$) than the inflationary one, if the tensor tilt $n_T\sim 1$. The overall spectrum, nonetheless, can exhibit the features of cosmic string-radiated  GW ($\Omega_{\rm GW}^{\rm CS}(f_{\rm dip})>\Omega_{\rm GW}^{\rm BGW}(f_{\rm dip})$--a plateau in the middle) for small values of $n_T$. Although the scenario then is disfavoured by the current PTA data.

\vspace{0.5cm}
\noindent
\section{Summary and conclusion}
\label{sec5}

Several PTA experiments, NANOGrav, EPTA+InPTA, PPTA as well as CPTA have reported strong evidence for a stochastic common spectrum process with  Hellings-Downs inter-pulsar correlations, suggesting a possible breakthrough towards the detection of a stochastic gravitational waves (GW) background at nano-Hz frequencies. While GW from SMBHB is not ruled out, another viable possibility namely, GW of cosmological origin (e.g., blue-tilted inflationary gravitational wave spectrum) provides an excellent fit to the recent data, e.g., to the NANOGrav 2023 data \cite{NANOGrav:2023hvm}. Inflationary GW with large blue-tilt not only generate a strong signal at nano-Hz frequencies, they offer the luring possibility to test post-inflationary cosmology with characteristic spectral features at higher frequencies. Generally, such blue-tilted GW saturate BBN bound on $N_{\rm eff}$, disallowing high $T_{\rm RH}$ temperature. However, if an entropy production epoch follows the standard reheating, BGW can evade the BBN constraints even for large $T_{\rm RH}$. Besides, such a post-inflationary scenario also creates unique spectral features testable at multiple detectors spanning a wide range of frequencies. We show that the Miracle-less WIMP DM model \cite{Borah:2022byb} naturally requires a dark matter mass-dependent matter epoch leading to entropy production prior to the BBN, and imprints blue-tilted GW. DM mass in the MeV-GeV ballpark makes inflationary GW compatible with NANOGrav and generates another peak testable with the next LIGO runs. Because of their weak interaction cross-section, Miracle-less WIMPs naturally explain null results in dark matter direct detection, and standouts as one of the few, perhaps, the only WIMP dark matter candidate so far, offering a PTA-LIGO complementarity.

\acknowledgements
The work of D.B. is supported by the science and engineering research board (SERB), Government of India grant MTR/2022/000575. R. S. is supported by the MSCA-IF IV FZU - CZ.02.2.69/0.0/0.0/20 079/0017754 project
and acknowledges European Structural and Investment Fund and the Czech Ministry of
Education, Youth and Sports.
\bibliographystyle{apsrev}
\bibliography{ref1,ref2,ref3}

\end{document}